\documentclass[a4paper]{jpconf}
\usepackage{graphicx}

\newcommand{\bfpsi}{\mbox{\boldmath$\psi$}}

\begin{document}

\title{Low temperature magnetic structure of the quasi 1-dimensional magnet Ni$_2$SiO$_4$}

\author{R H Colman$^1$, T Fennell$^{1,2}$, C Ritter$^2$, G Lau$^3$, R J Cava$^3$ and A S Wills$^1$}

\address{$^1$University College London, 20 Gordon Street, London, WC1H 0AJ, UK \\$^2$Institut Laue-Langevin, BP 156, 38042 Grenoble Cedex 9, France \\$^3$Princeton Materials Institute and Department of Chemistry, Princeton University, Princeton, New Jersey, USA}

\ead{a.s.wills@ucl.ac.uk}

\begin{abstract} Ni$_2$SiO$_4$, Liebenbergite, is an example of a quasi-one-dimensional magnet made up of frustrated corner sharing triangles of Ni$^{2+}$ ($S = 1$) ions that propagate parallel to the {\it b} axis. Ni$_2$SiO$_4$ is isostructural with olivine, a common mineral of varying composition Fe$_{2-x}$Mg$_{x}$SiO$_4$, and is described in the orthorhombic space group {\it Pnma}. A synthetic polycrystalline sample of Ni$_2$SiO$_4$ was studied using constant wavelength powder neutron diffraction. Diffraction spectra were collected above and below the antiferromagnetic ordering transition ($T_{\textrm{\tiny{N}}}$ $\approx$ 34\,K) and were used to refine the atomic and magnetic structures of Ni$_2$SiO$_4$. Corepresentational theory was used to determine the symmetry-allowed magnetic structures after the N\'eel transition and the refined magnetic structure evidences both ferromagnetic and antiferromagnetic inter-chain interactions, and ferromagnetic intra-chain coupling. The competition between the magnetic interactions can be seen in the canting of the moments away from a collinear arrangement. 
\end{abstract}


Current theories of magnetism are challenged by highly correlated systems that show degeneracies and fluctuations. The characterisation of ground states of new model systems is an important step for testing and confirming predictions, and geometrically frustrated magnets provide important opportunity for this. A triangle of antiferromagnetically coupled spins is the archetypal example of a geometrically frustrated motif, due to the competing exchange interactions between the three moments. There are a huge variety of ways to build these triangles into higher dimensional structures, with particular interest in structures with low connectivity, such as the vertex sharing triangle and tetrahedral geometries of the kagom\'e \cite{kagome} and pyrochlore \cite{pyrochlore} lattices respectively. 

\begin{figure}[t]
\begin{center}
\includegraphics[scale=0.25]{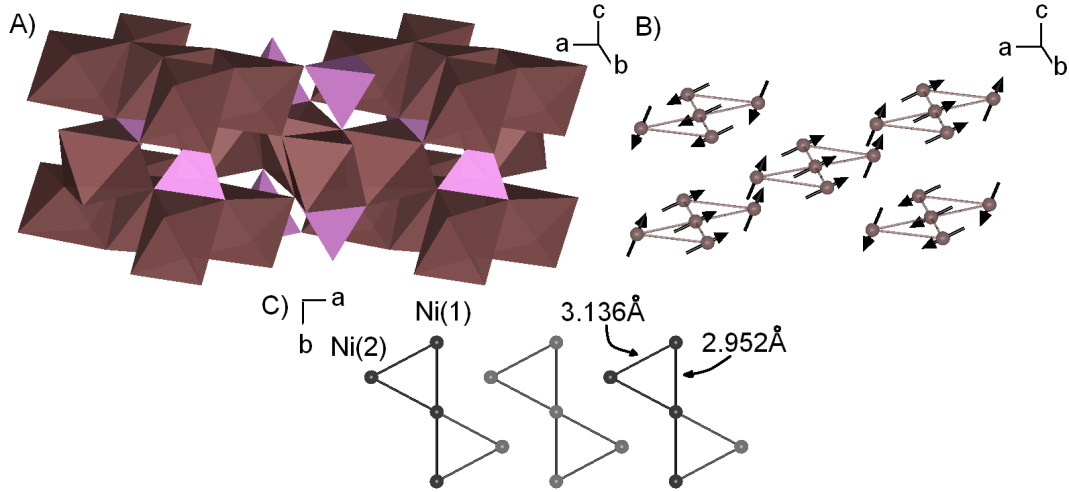}
\end{center}
\caption{\label{structure}{\bf A}: A polyhedral representation of Ni$_2$SiO$_4$ with NiO$_6$ displayed as dark edge-sharing octahedra and SiO$_4$ displayed as light tetrahedra occupying the interstitial sites between the chains; {\bf B}:   The magnetic structure of Ni$_2$SiO$_4$ refined from neutron diffraction data collected at 2\,K and $\lambda = 1.91$\,\AA \  showing the zig-zag chain arrangement of corner sharing isosceles triangles of Ni$^{2+}$ ions propagating along the $b$-axis; {\bf C}: A view parallel to the $b$-axis giving the edge-lengths of the isosceles triangles formed from the Ni$^{2+}$ ions (Generated using FPStudio \cite{fpstudio}).}
\end{figure}

The olivine family of minerals ($XY$O$_4$ where $X$ = an M$^{2+}$ transition metal ion, often with mixed occupancy and $Y$ = a group III or group IV atom, most commonly Si) consist of a one dimensional array of metal ions which form zig-zag chains of corner-sharing triangles. In Ni$_2$SiO$_4$ there are two symmetry distinct metal sites: Ni(1) form linear chains of ions propagating parallel to the {\it b}-axis, and Ni(2) make up the apex of the triangles zig-zagging along the linear chains within the $ab$-plane (figure~\ref{structure}B). These metal sites are coordinated by an octahedra of oxygen, and the tetrahedral interstitial sites between the chains are occupied by group III or group IV atoms (figure~\ref{structure}A).

Preliminary inelastic powder neutron studies of this material show the presence of a spin gap at $|Q|\approx 0.8$\,\AA$^{-1}$ that opens below T$=35$\,K at the transition to the ordered state.  This gapped behaviour was first eluded to by inverse susceptibility measurements that were well fitted to a spin gap plus Curie-Weiss model\cite{cava} that was predicted as a characteristic of a one dimensional chain of integer spin\cite{haldane}. This is in contrast to the gapless behaviour predicted and observed for chains of non-integer spin.

Diffraction spectra from a 15\,g sample of Ni$_2$SiO$_4$ \cite{cava}, were recorded on the D1A powder neutron diffractometer at the ILL, both above and below the N\'eel transition using neutrons of wavelength $\lambda = 1.91$\,\AA. The sample temperature was controlled using a standard `orange' cryostat and the sample was contained within a vanadium can. The refinement of the collected spectra at 39\,K and 2\,K will be presented.  
 


The refined nuclear structure of Ni$_2$SiO$_4$ at 39\,K was consistent with previously published powder diffraction results \cite{cava} and the low temperature diffraction pattern there were new magnetic Bragg peaks that could be indexed with a magnetic propogation vector {\bf k$_{24}$}=($\frac{1}{2}, 0, \frac{1}{2}$) in Kovalev's notation \cite{kovalev}, also consistent with previously published results \cite{newnham}, although the refined magnetic structure was found to differ from that reported. The different symmetry-types of magnetic structure were determined using corepresentation analysis and the program SARA{\it h} \cite{sarah}. Refinement of the magnetic structure was performed in terms of the basis vectors, $\psi_\nu$, of a given  irreducible corepresentation multiplied by a (weighting) coefficient, $C_\nu$. In this formalism the set of refined moments, $m_j$, is generated from the linear combination of the basis vectors: $m_j=\sum_\nu C_\nu\psi_\nu$.

\begin{figure}[t]
\begin{center}
\includegraphics[scale=0.6]{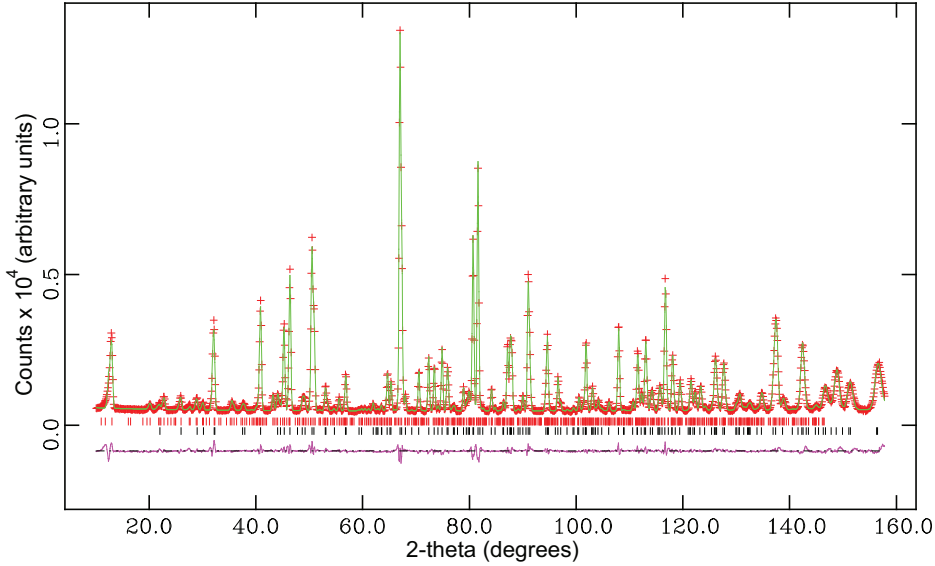}
\end{center}
\caption{\label{pattern}The observed and calculated neutron diffraction spectrum of Ni$_2$SiO$_4$ with nuclear and magnetic reflections, collected on D1A of the ILL at 2\,K using neutrons of $\lambda = 1.91$\,\AA \cite{gsas}. The crosses correspond to experimental data; the lines correspond to calculated values and the upper and lower tick-marks indicate the positions of the magnetic and crystallographic reflections respectively. The goodness of fit parameters were $\chi^2 = 12.52$ and $R_{wp} = 0.0461$ for 35 parameters.}
\end{figure} 

The data could only be well fitted using a corepresentation formed by the combination of the representations $\Gamma_3$ and $\Gamma_7$ from Kovalev's tables (table~\ref{bvtable}) \cite{kovalev,zoso}. It is notable that the 4 equivalent positions  of each of the Ni crystallographic sites form distinct groups, or {\it orbits}, formed from positions (1+3) and (2+4) and the combination of these representations defines moment components on all of the symmetry related positions. The equivalence of these orbits prevents the sizes of the moments from being separately refined, and those moments of a given Ni site were constrained to be equal in magnitude.  The refinement (figure~\ref{structure}C) was not improved by the presence of  components parallel to the $b$-axis, and the moments of both the Ni(1) and Ni(2) sites were subsequently restricted to the $ac$-plane. The inter-chain couplings are shown to be antiferromagnetic when along the $<-1~0~1>$ direction and ferromagnetic along the $<1~0~1>$ direction, with ferromagnetic interactions dominating between the intra-chain spins. The moments on the two sites form a non-collinear structure, with the canting of Ni(1) and Ni(2) moments away from the $a$-axis being $26.26^{\circ}$ and $69.42^{\circ}$ respectively. The moments of the Ni(1) and Ni(2) sites refined to  2.03(3) and 2.08(4) $\mu_B$, respectively, in good agreement with expectations for $S=1$ Ni$^{2+}$. The refined structure differs from that previously reported in two respects.\cite{newnham} 1.) No mention of an attempt to refine the moments of the two crystallographically distinct Ni$^{2+}$ sites to different values was made but we found that an improvement in the refinement could be achieved if these values were allowed to refine separately. 2.) Symmetry analysis shows that $\Gamma_1$ and $\Gamma_5$ would need to be used to describe the previously reported structure, in contrast to $\Gamma_3$ and $\Gamma_7$ from our results. The differences in the diffraction pattern between these two structures are subtle but our goodness of fit was noticeably reduced if the previously reported structure was used to model the data. The subtle differences could easily have been masked by the resolution of the previously reported pattern along with the limited angular range reported (5 - 30 2$\theta$) in comparison to our high resolution data collected (5 - 160 2$\theta$).

\begin{table}[h]
\begin{center}
\begin{tabular}{ccc|ccc|ccc}
  IR  &  BV  &  Atom & \multicolumn{6}{c}{BV components}\\
  & & & \multicolumn{3}{c}{Ni(1)} & \multicolumn{3}{c}{Ni(2)} \\
      &      &             &$m_{\|a}$ & $m_{\|b}$ & $m_{\|c}$ &$m_{\|a}$ & $m_{\|b}$ & $m_{\|c}$ \\
\hline
$\Gamma_{3}$ & $\bfpsi_{1}$ &      1 &      1 &      0 &      0 &      1 &      0 &      0   \\
             &              &      2 &      0 &      0 &      0 &      0 &      0 &      0   \\
             &              &      3 &      1 &      0 &      0 &      1 &      0 &      0   \\
             &              &      4 &      0 &      0 &      0 &      0 &      0 &      0   \\
             & $\bfpsi_{2}$ &      1 &      0 &      1 &      0 &      0 &      0 &      1   \\
             &              &      2 &      0 &      0 &      0 &      0 &      0 &      0   \\
             &              &      3 &      0 &     -1 &      0 &      0 &      0 &      1   \\
             &              &      4 &      0 &      0 &      0 &      0 &      0 &      0   \\
             & $\bfpsi_{3}$ &      1 &      0 &      0 &      1 &       &       &        \\
             &              &      2 &      0 &      0 &      0 &       &       &        \\
             &              &      3 &      0 &      0 &      1 &       &       &        \\
             &              &      4 &      0 &      0 &      0 &       &       &        \\
$\Gamma_{7}$ & $\bfpsi_{1}$ &      1 &      0 &      0 &      0 &       0 &      0 &      0  \\
             &              &      2 &      1 &      0 &      0 &      -1 &      0 &      0  \\
             &              &      3 &      0 &      0 &      0 &       0 &      0 &      0  \\
             &              &      4 &      1 &      0 &      0 &      -1 &      0 &      0  \\
             & $\bfpsi_{2}$ &      1 &      0 &      0 &      0 &       0 &      0 &      0  \\
             &              &      2 &      0 &     -1 &      0 &       0 &      0 &      1  \\
             &              &      3 &      0 &      0 &      0 &       0 &      0 &      0  \\
             &              &      4 &      0 &      1 &      0 &       0 &      0 &      1  \\
             & $\bfpsi_{3}$ &      1 &      0 &      0 &      0 &       &       &        \\
             &              &      2 &      0 &      0 &     -1 &       &       &        \\
             &              &      3 &      0 &      0 &      0 &       &       &        \\
             &              &      4 &      0 &      0 &     -1 &       &       &        \\
\end{tabular}
\caption{Basis vectors for representations $\Gamma_3$ and $\Gamma_7$ of the space group $Pnma$ with {\bf k}$_{24}$=($\frac{1}{2}, 0, \frac{1}{2}$) following Kovalev's notation\cite{kovalev}. The equivalent positions of  Ni(1) are defined according to 1: $( 0,~ 0,~ 0)$, 2: $( .5,~ .5,~ .5)$, 3: $( 0,~ .5,~ 0)$, 4: $( .5,~ 0,~ .5)$; and for Ni(2) according to 1: $( .27404,~ .25,~ .99123)$, 2: $( .77404,~ .25,~ .50877)$, 3: $( .72596,~ .75,~ .00877)$, 4: $( .22596,~ .75,~ .49123)$}
\label{bvtable}
\end{center}
\end{table}


In conclusion, we report a low temperature powder neutron diffraction study of Ni$_2$SiO$_4$, a quasi-1-dimensional magnet made up of zig-zagging chains of Ni$^{2+}$ triangles. The magnetic structure is shown to reveal evidence of frustrated interactions with a significant canting away from collinearity. We note also that corepresentational theory is required to fully describe the symmetry of the spin configuration which we found to have subtle differences to that previously published.

\section{Acknowledgments}
We would like to thank Clemens Ritter for his invaluable help with data collection and the EPSRC  (grant number EP/C534654)  for funding.

\end{document}